\documentclass[prl,aps,amssymb,showpacs,twocolumn]{revtex4}
\usepackage{amsmath}
\usepackage{amssymb}
\usepackage{amsthm}
\usepackage{amsfonts}
\usepackage{enumerate}
\usepackage{latexsym}
\usepackage{graphicx}

\newcommand{\beq}{\begin{equation}}
\newcommand{\eneq}{\end{equation}}

\input{epsf}

\begin{document}

\tolerance 10000

\newcommand{\vk}{{\bf k}}


\title{Properties of Non-Abelian Fractional Quantum Hall States at Filling $\nu=\frac{k}{r}$ }

\author{B. Andrei Bernevig$^{1,2,3}$ and F.D.M. Haldane$^2$}

 \affiliation{$^1$Princeton Center for Theoretical
Physics, Princeton, NJ 08544} \affiliation{$^2$ Department of
Physics, Princeton University, Princeton, NJ 08544}
\affiliation{$^3$ Department of Physics, University of California, Irvine, CA 92697}

\begin{abstract}
We compute the physical properties of non-Abelian Fractional Quantum Hall (FQH) states described by Jack polynomials at general filling $\nu=\frac{k}{r}$. For $r=2$, these states are identical to the $Z_k$ Read-Rezayi parafermions, whereas for $r>2$ they represent new FQH states. The $r=k+1$ states, multiplied by a Vandermonde determinant, are a non-Abelian alternative construction of states at fermionic filling $2/5, 3/7, 4/9...$. We obtain the thermal Hall coefficient, the quantum dimensions, the electron scaling exponent, and show that the non-Abelian quasihole has a well-defined propagator falling off with the distance. The clustering properties of the Jack polynomials, provide a strong indication that the states with $r>2$ can be obtained as correlators of fields of non-unitary conformal field theories, but the CFT-FQH connection fails when invoked to compute physical properties such as thermal Hall coefficient or, more importantly, the quasihole propagator. The quasihole wavefuntion, when written as a coherent state representation of Jack polynomials, has an identical structure for all non-Abelian states at filling $\nu=\frac{k}{r}$.
\end{abstract}

\date{\today}


\maketitle

The proposals for non-Abelian FQH spin-polarized states have so far been largely confined to the Read-Rezayi sequence \cite{read1999}, which expresses the bulk trial wavefunctions as correlators of fields of the $Z_k$ parafermion conformal field theories \cite{zamolodchikov1985}. The properties of these states, such as the dimension of the Hilbert space of quasihole excitations and the braiding properties, are obtained by using complicated CFT machinery \cite{ardonne2002,read1999}. In \cite{bernevig2007} we have obtained the Read-Rezayi wavefunctions without appealing to CFT: we showed that they are identical to a series of polynomials known in mathematics as Jack polynomials (Jacks). The Jacks also provide for an infinite series of non-Abelian FQH states beyond the Read-Rezayi sequence. These states (we find) are related to non-unitary CFTs, which some authors argue can never describe physical FQH states \cite{read2007}. However, as trial wavefunctions, they seem essentially similar to the states related to unitary CFTs, so we see no reason at this time to reject them a priori. Some of these states could provide for alternative non-Abelian constructions of trial wavefunctions at the experimentally important fillings of $2/5, 3/7, 4/9...$. In fact, exact diagonalization and overlap finite-size studies show that some of these states are competitive with the Abelian Jain states at the same fillings \cite{simon2006}. We now understand this competitiveness analytically \cite{bernevig2008}, as the non-Abelian state at, for example, $\nu=\frac{2}{5}$ \cite{simon2006} is remarkably close to the Jain state at the same filling for small number of electrons.

This paper is devoted to computing the physical properties of the FQH states introduced in \cite{bernevig2007}. The Jacks describe not only the uniform ground-state but also all droplet states swelled by the introduction of quasiholes, and hence describe the full family of gapless edge states of the droplet. We analytically compute the thermal Hall (Leduc-Reghi) coefficient $c$, and, numerically, the electron propagator exponent $g_e$. We give analytic expressions for the dimension of the quasihole subspace. We derive an expression for the fractionalized pinned quasihole wavefunction which puts all FQH states on the same footing. We provide strong evidence that the Jacks can be related to the correlation functions of $W_k(k+1,k+r)$ CFTs, as was initially conjectured in \cite{feigin2002}.
The CFT-FQH correspondence breaks down when calculating the quasihole propagator $g_{qh}$ on the edge: for the $r>2$ Jacks, the CFTs are non-unitary and would predict a negative exponent. These predictions are based, at their core, on an unproven postulate: that properties computed using the CFT machinery, most importantly, the CFT scalar product, are identical to properties computed using the quantum mechanical scalar product in two space dimensions. We show that this postulate, conjectured to be valid for unitary CFTs, must break down for non-unitary CFTs: although we could not compute the exact value of the non-Abelian quasihole propagator exponent, we present numerical proof that it is a well defined positive number.

\begin{figure}
\includegraphics[width=3.3in, height=1.9in]{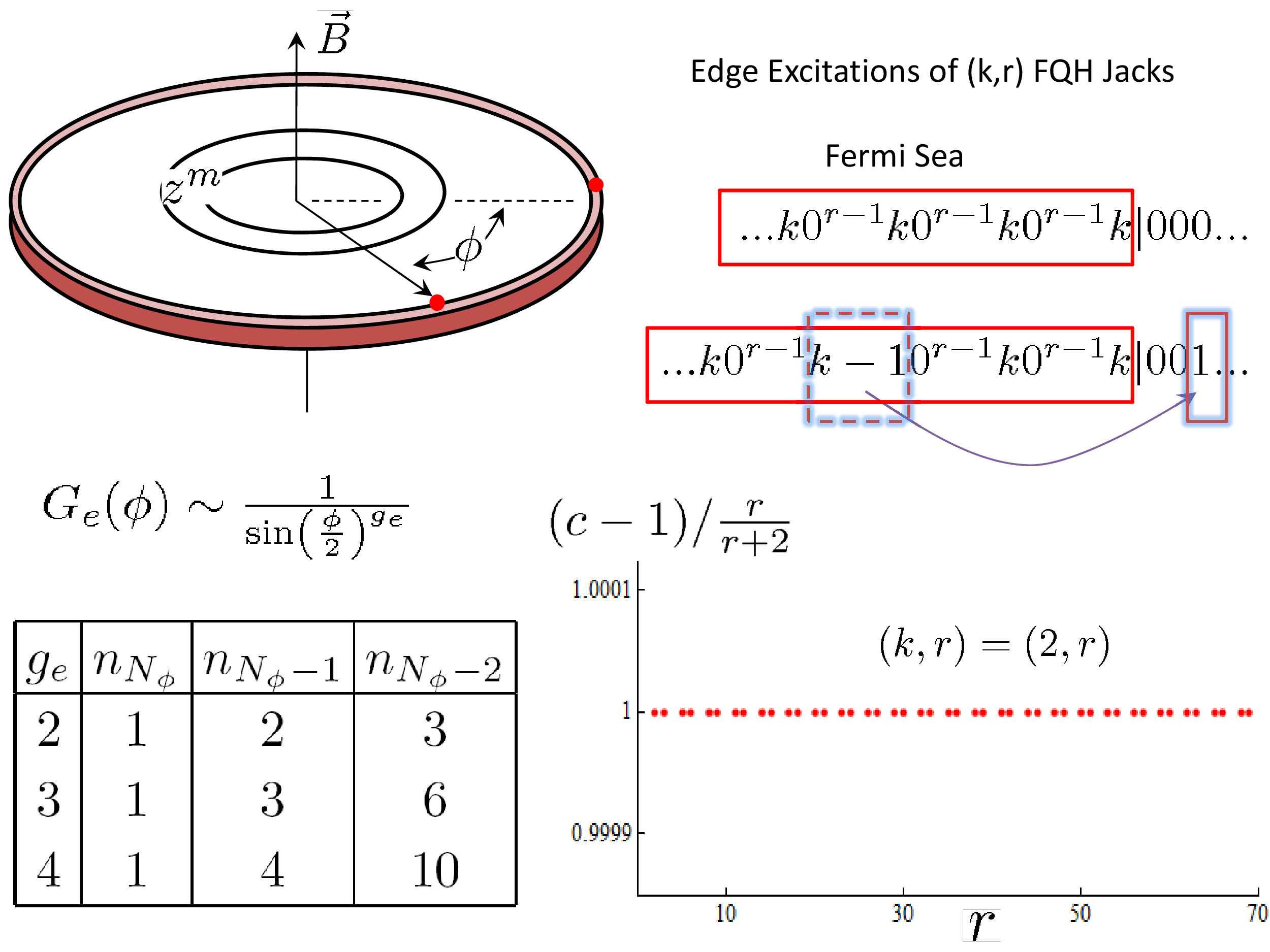}
\caption{Left: FQH problem on a disc has the electron propagator on the edge depending on the chord angle through a power-law. The occupation number of the angular momentum orbitals is then an imprint of the state which gives the electron scaling exponent $g_e$. Upper Right: The thermal Hall coefficient (edge specific heat) can be determined by counting the number of particle-hole excitations above the Fermi sea on the edge. Lower Right: Plot of the specific heat for $(k,r)=(2,r)$ FQH Jack state. We find $c= \frac{r}{r+2}$. Other cases were treated analytically in the text.}\label{edgephysics}
\end{figure}

Jacks $J^{\alpha}_{\lambda}(z)$ are symmetric polynomials in $z$
$\equiv$ $\{z_1,z_2,\ldots ,z_N\}$, labeled by a partition $\lambda$
with length $\ell_{\lambda} \le N$, and a parameter $\alpha$;
$\lambda$ can be represented as a (bosonic) occupation-number
configuration $n(\lambda)$ = $\{n_m(\lambda),m=0,1,2,\ldots\}$ of
each of the lowest Landau level (LLL) orbitals $\phi_m(z) = (2\pi m! 2^m)^{-1/2} z^m \exp(-|z|^2/4)$ with angular momentum
$L_z = m \hbar$ (see Fig[\ref{edgephysics}]), where, for $m > 0$,
$n_m(\lambda)$ is the multiplicity of $m$ in $\lambda$. When
$\alpha$ $\rightarrow$ $\infty$, $J^{\alpha}_{\lambda}$
$\rightarrow$ $m_{\lambda}$, which is the monomial wavefunction of
the free boson state with occupation-number configuration
$n(\lambda)$; a key property of the Jack $J^{\alpha}_{\lambda}$  is
that its expansion in terms of monomials only contains terms
$m_{\mu}$ where $\mu$ $\le$ $\lambda$, where $\mu$ $<$ $\lambda$
means the partition $\mu$ is \textit{dominated} by
$\lambda$\cite{stanley1989}. The ``dominance rule'' is identical with the ``squeezing
rule''\cite{sutherland1971} that connects configurations
$n(\lambda)$ $\rightarrow$ $n(\mu)$: ``squeezing'' is a two-particle
operation that moves a particle from orbital $m_1$ to $m_1'$ and
another from $m_2$ to $m_2'$, where $m_1 < m_1' \le m_2' < m_2$, and
$m_1+m_2$ = $m_1'+m_2'$; $\lambda > \mu$ if $n(\mu)$ can be derived
from $n(\lambda)$ by a sequence of ``squeezings''. Jacks of $N$ particles are also eigenstates of a set of $N$ self-commuting operators, including the angular momentum and the
Laplace-Beltrami operator $\mathcal H_{\rm LB}(\alpha)$ \cite{stanley1989}.

  The Fock space of our FQH states is spanned by Jacks uniquely described by two integers $(k,r)$, not necessarily relatively prime. These integers provide the filling factor $\nu=\frac{k}{r}$, the Jack polynomial parameter $\alpha(k,r) =-\frac{k+1}{r-1}$, the root partition (orbital occupation number) for the ground state $k0^{r-1}k0^{r-1}...k0^{r-1}k$, and the physical requirement that the Jack polynomial defined in this way vanishes when $k+1$ particles come together as the $r$'th power of the difference between their coordinates. The root partition is reminiscent of the orbital density-wave exhibited by the thin-torus limit (the non-interacting Tao-Thouless states \cite{tao1983}), but in fact, has little to do with these states, as the Jacks represent the faithful many-body interacting wavefunctions. The Read-Rezayi sequence is $r=2$, the Laughlin $\nu =\frac{1}{r}$ states are $k=1$ and the infinite remaining series of FQH states is, with several exceptions, new. In \cite{bernevig2007} \cite{bernevig2007B} we have showed that each FQH Fock space (groundstate and quasihole wavefunctions) is uniquely defined by a generalized Pauli principle that does not allow more than $k$ particles to be placed in $r$ consecutive angular momentum orbitals. In partition language, this means that the Fock space is spanned by Jack polynomials of $(k,r)$ admissible partitions $\lambda $ \cite{feigin2002}: $\lambda_i - \lambda_{i+k} \ge r$. To count the number of FQH states (unpinned quasiholes) of $N$ particles, at $n_\Phi$ fluxes added above the ground state, with angular momentum $l_z$, we have to count the number of partitions of $N$ parts $\lambda=(\lambda_1,...,\lambda_N)$, with $\lambda_1 \le N_\Phi=\frac{r}{k}(N-k) + n_\Phi$, with $\sum_{i=1}^N \lambda_i = l_z+ \frac{1}{2}N N_\Phi$, satisfying the Pauli principle $\lambda_i - \lambda_{i+k} \ge r$. The weight $\sum_{i=1}^N \lambda_i$ equals the homogeneous dimension of the polynomial. Since expressions for the number of partitions satisfying these properties have been obtained in the theory of jagged partitions, the counting of the quasihole states becomes a simple matter. In the $r=2$ case, this becomes identical to the formulas obtained using CFT methods in \cite{ardonne2002}. Using this counting, we now obtain the values of the thermal Hall coefficient in the cases $(k,2)$, $(k,3)$ and $(2,r)$.

In FQH states, the thermal Hall coefficient is identical to the edge specific heat, which can be obtained by computing the entropy of the edge states. In the high-temperature expansion $\beta/L <<1$, Cardy \cite{cardy1988} showed that for a CFT, the partition function scales as $\exp( \frac{L \pi T}{6 } c)$ with $c$ a rational number related to the central charge of the CFT. We assume a circular FQH gapless edge of length $L$ and create particle-hole excitations of energy $v_F (k- k_F) = \frac{2 \pi v_F}{L} (D-D_F) $ where $D- D_F = D- \frac{k}{2r }N(N-k)$ is the angular momentum $l_z$ of the edge excitation and $v_F$ is a non-universal Fermi velocity whose value depends on the details of the edge potential. The number of these excitations $N^{(k,r)}_{(N,D)}$, at each $l_z$,  is given by the number of all $(k,r)$ admissible partitions $N$ of weight $\sum_{i=1}^N \lambda_i= D$, with no restriction on $\lambda_1$. For the case of parafermions - $(k,r)= (k,2)$ - , this number is easily derived in Andrews' book \cite{andrews1998} on the theory of partitions in terms of the generating function
\begin{equation} \label{generatingfunction}
p(q,z)=\sum_{n_1..n_k=0}^{\infty} \prod_{i=1}^k \frac{q^{ N_i^2} (q z)^{ N_i} }{(q)_{n_i}}; \;\;\; N_i = \sum_{j=i}^k n_j
\end{equation}
where $(q)_m = \prod_{i=1}^m (1- q^i)$. The number of $(k,2)$ Jacks of $N$ particles with total homogeneous dimension $D$ is then $\#^{(k,2)}_{(N,D)} =\frac{1}{N! D!} \frac{\partial^N \partial^D}{\partial z^N \partial q^D} p(q,z)\mid_{z,q =0}$. The partition function of the edge excitations in the canonical enssamble is then $Z =  \sum_{D=0}^\infty \#^{(k,2)}_{(N,D)} \exp(- \frac{2 \pi v_F \beta}{L} (D-D_F))$ with $\beta =1/T$. In the thermodynamic limit $N\rightarrow \infty$, a $U(1)$ decoupling occurs:
\begin{equation} \label{partitionfunction}
Z(T) =\frac{q^{\frac{N^2}{2k}}}{(q)_{\infty}} \sum_{n_1..n_{k-1}} \frac{q^{\frac{1}{2} n^T B n}}{(q)_{n_1}..(q)_{n_{k-1}}};\;\;\; q=e^{-\frac{2\pi v_F \beta}{L}}
\end{equation}
where $n^T =(n_1,...n_{k-1})$, and the symmetric matrix $B$ reads $B_{ij} = \frac{2}{k}i(k-j)$ for $i\le j$. Several other terms, all linear in $n$, usually appear in the numerator of the summand, but they are irrelevant for the asymptotic scaling with temperature and we disregard them. The edge partition function has decoupled in a product of two factors (this occurs for all the $(k,r)$ states studied here): we recognize the $\frac{q^{\frac{N^2}{2k}}}{(q)_{\infty}}$ as the abelian $U(1)$ part of the excitation spectrum (with  $\frac{N^2}{2k}$ representing the infinite, ground-state contribution). This gives a specific heat $c=1$, which adds to the non-Abelian part. The latter can be obtained from our partition function by saddle-point evaluation of its asymptotic form, closely following Nahm \emph{et. al.} \cite{nahm1993} and  Richmond and Szekeres \cite{richmond1981}:
\begin{equation} \label{centralchargesaddlepoint}
c =1+ \frac{6}{\pi^2} \sum_{i=1}^{k-1} {\cal{L}}\left(\frac{1}{d_i^2}\right);\;\;\;\; d_i = \prod_{j=1}^{k-1} \left(\frac{d_j}{\sqrt{d_j^2-1}} \right)^{B_{ij}}
\end{equation}
\noindent where ${\cal{L}} = \sum_{n=1}^\infty \frac{z^n}{n^2} + \frac{1}{2} \ln(z) \ln(1-z)$ is the Rogers dilogarithm function, and the $d_i$ are the solutions of the system of equations above. For the $(k,r)=(k,2)$ case, from the second part of Eq.(\ref{centralchargesaddlepoint}), we find that the $d_i$'s satisfy the recursion equations of Chebyshev's polynomials of the second kind, which in turn fixes them to be the first column of the modular $S$-matrix (the quantum dimensions) of the $SU(2)_k$ WZW model:
\begin{equation}
d_i^2-1 = d_{i-1} d_{i+1};\;\; d_0=d_{k} =1;\;\; d_i =\frac{\sin \frac{(i+1)\pi}{k+2} }{\sin \frac{\pi}{k+2}}
\end{equation}
\noindent An almost identical result was obtained by Nahm and collaborators \cite{nahm1993} starting from the CFT character formulas, but in their case, the summation in Eq.(\ref{centralchargesaddlepoint}) is limited to $[k/2]$ and the resulting central charge is that of the $M(2,2+k)$ minimal Virasoro model and not of the $SU(2)_k$ WZW models. Using the dilogarithm identities found by Kirilov \cite{kirilov1995}, we then find $c=1+ \frac{2(k-1)}{k+2}$. We have hence obtained both the known quantum dimensions and the central charge of the $Z_k$ parafermion FQH states without using CFT but making use of partition counting and the generalized Pauli principle.

We now turn to the derivation of the thermal Hall coefficient for the FQH states (Jacks) described by the $(k,r)=(k,3)$ generalized Pauli principle. The generating function (equivalent of Eq.(\ref{generatingfunction})) for the number of partitions $\lambda_i - \lambda_{i+k} \ge 3$ of $N$ parts with $\sum_{i=1}^N \lambda_i = D$ has been obtained using the theory of jagged partitions \cite{fortin2005} \cite{feigin2002B}. After some algebra, we find the same partition function as in Eq.(\ref{partitionfunction}) but with ground-state energy $3\frac{N^2}{2k}$ and the matrix $B_{ij}$:
\begin{equation}
B_{ij} = 2 \min(i,j) + \max(0, i+j - k) - \frac{3}{k} i\cdot j
\end{equation}
\noindent where $i,j =1...k-1$. Then $c$ takes the same form as in Eq.(\ref{centralchargesaddlepoint}) but with the $d_i$'s satisfying a modified recursion:
\begin{equation} \label{secondrecursion}
d_i \sqrt{d_i^2-1} = d_{i+1} d_{i-1};\;\; d_0 =1,\;\;d_{\frac{k}{2}+i} = d_{\frac{k}{2}-i}
\end{equation}
\noindent for $i=0...\frac{k}{2}$ (since $r=3$ and $k+1$, $r-1$ are coprime in the Jacks, it follows that $k$ is even; however, everything below is also valid for $k$ odd, with $\frac{k}{2} \rightarrow \left[\frac{k}{2}\right]$). The quantum dimensions $d_i$ in relation Eq.(\ref{secondrecursion}) are the zeroes of a set of polynomials that generalize the Chebyshev polynomials found in the case $(k,r)=(k,2)$. However, we find that these polynomials cannot (unlike Chebyshev's) be obtained as the limit of any Jacobi polynomials, and hence conjecture that they are new, and that they define a new set of dilogarithm identities. Further study of their properties is necessary. By explicitly computing the sum over dilogarithms, we find that the edge specific heat for the FQH states (Jacks) at $(k,r)=(k,3)$ is:
\begin{equation}
c =1+ \frac{6}{\pi^2} \sum_{i=1}^{k-1} {\cal{L}}\left(\frac{1}{d_i^2}\right) =1+ \frac{3 (k-1)}{k+3}
\end{equation}
\noindent

  Finally, we turn to the derivation of the thermal Hall coefficient for the FQH states (Jacks) described by the $(k,r)=(2,r)$ generalized Pauli principle. Since $k=2$, they are all paired states. The $r=2$ state is the Pfaffian state and the $r=3$ state is the new Gaffnian state\cite{simon2006,bernevig2007}. The generating function (equivalent of Eq.(\ref{generatingfunction})) for the number of $(2,r)$ partitions $\lambda_i - \lambda_{i+2} \ge r$ has been obtained using the theory of jagged partitions \cite{fortin2005B} \cite{feigin2002C}. After some algebra, we find a ground-state energy $r\frac{N^2}{4}$ and a matrix $B_{ij}$:
\begin{equation}
B_{ij} =(s-\frac{r}{4}) \delta_{i,1}\delta_{j,1}+ (s+1-\max(i,j))(1-\frac{\delta_{i,1}+\delta_{j,1}}{2}) \nonumber
\end{equation}
\noindent Although we were not able in this case to find a particularly elegant analytic form for the recursion relation satisfied by the $d_i$'s, we have solved the equations numerically (and analytically for small values of $r$), and we find (see Fig[\ref{edgephysics}]), for $r-1$ not divisible by $k+1=3$, that $c=\frac{r}{r+2}$. We are now in a position to form a strong conjecture. We have analytically found that $c=1+\frac{2(k-1)}{k+2}$ for $(k,r)=(k,2)$, $c=1+\frac{3(k-1)}{k+3}$ for $(k,r)=(k,3)$, $c=1+\frac{r}{2+r}$ for $(k,r)=(2,r)$, and $c=1$ for $(k,r) = (1,r)$, the latter ones representing the $\nu=\frac{1}{r}$ Laughlin states. We hence conjecture that the thermal Hall coefficient for the FQH states (Jacks) described by the $(k,r)$ generalized Pauli principle is:
\begin{equation}
c =1+ \frac{r(k-1)}{k+r}
\end{equation}

We now numerically compute the electron propagator exponent $g_e$ on the edge of the $(k,r)$ FQH state. Following Wen \cite{wen1992}, we let the equal-time propagator
exponent for the electron on the edge (radius $R = \sqrt{2 r N/k}$) of the $N$ particle FQH droplet be $G(\phi) \sim \exp(i (N_\Phi+ g_e/2) \phi) /\sin(\phi/2)^{g_e}$.  $\phi$ is the arc angle and $N_\Phi$ is the number of magnetic fluxes. The Fourier transform of this gives the occupation number $n_L$ of orbitals with angular momentum $L$:
\begin{equation}
n_L= \int_0^{2 \pi}  e^{-i L \phi} G(\phi) d \phi \sim \left(%
\begin{array}{c}
  N_\Phi-L+g_e-1 \\
  N_\Phi-L \\
\end{array}%
\right)
\end{equation}
for $L \le N_\phi$ and $n_L=0$ for $L>N_\phi$. For fermions, $g_e=1$ and one obtains the occupation of a zero temperature Fermi sea, whereas for $g_e>1$, as is the case for FQH states, we obtain the well known Luttinger Liquid behavior $n_L
\sim (N_\phi - L)^{g_e-1}$.
Since the constant of proportionality in front is the same, we can
compute the ratios:
$n_{N_\phi}:n_{N_\phi-1}:n_{N_\phi-2}...$. These patterns of integers then give the imprint of the
state (see Fig.[\ref{edgephysics}]). It follows that $n_{N_\phi-1}:n_{N_\phi}=g_e$ and $n_{N_\phi-2}:n_{N_\phi}=g_e(g_e+1)/2$. We numerically compute $n_L =\int d^2 z \int d^2 z' \phi_L(z)^\star n(z,z') \phi_L(z')$ where $n(z,z')$ is the one-body density matrix for the FQH in consideration, and $\phi_L(z)$ are the single-particle LL orbitals. For $(k,r)=(1,2)$ Laughlin state for $N=7$ particles we find $1.: 2.:
2.880271139$ which is close to  the
thermodynamic limit ratio of $1:2:3$. For $(k,r)=(1,4)$ Laughlin state for $N=6$ particles we find $1: 4:
9.480108534$ which is close to the expected $1:4:10$. From these we infer that $g_e=r$ for the $\nu=1/r$ Laughlin states, thereby reproducing a well-known result obtained by Wen \cite{wen1992}. For $(k,r)=(2,2)$ Read-Moore for $N=10$ particles we find
$1:2.211521156:3.435931591$ for $N=10$ particles, which is close to
$1:2:3$ ($g_e=2$). For $(k,r)=(4,3)$ and $N= 12$ particles we find $1:3.079004031:6.247432130$,  which is close to $1:3:6$ ($g_e=3$). For  $(k,r)=(3,4)$  and $N=9$ particles we find
$1:3.933251112:8.424840830$, which is close to
a  thermodynamic limit  of  $1:4:10$ ($g_e=4$). We have performed a series of numerical evaluations from which we see our data is consistent with the value
\begin{equation}
g_e=r
 \end{equation}
\noindent in the $(k,r)$ sequence. We later show that this is also supported by the clustering conditions of the Jack polynomials.

\begin{figure}
\includegraphics[width=3.3in, height=1.9in]{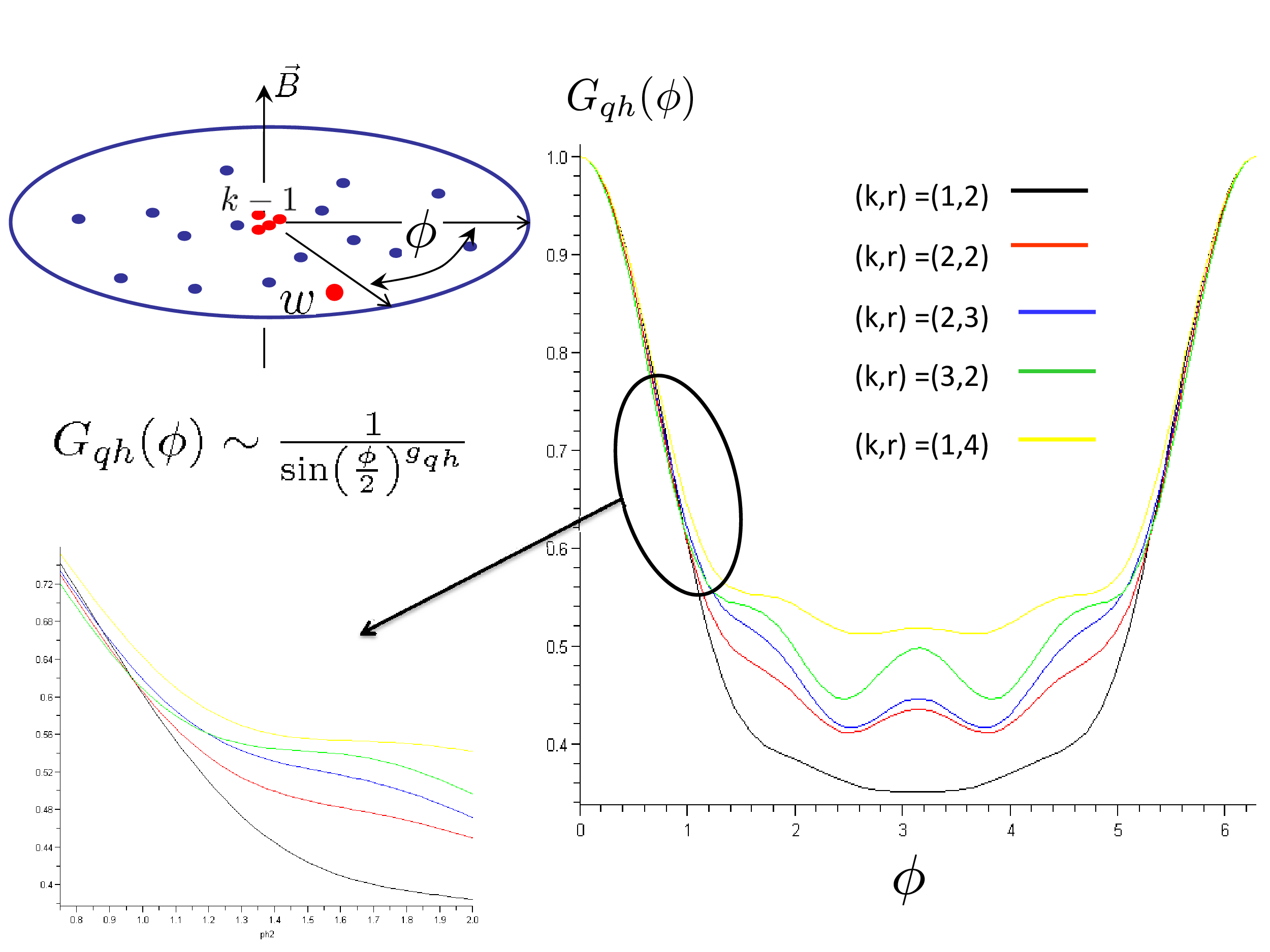}
\caption{ Finite-size quasihole propagator as a function of the arc distance on the edge of the FQH state. We place $k-1$ quasiholes at the origin in a disk geometry and leave one fractionalized quasihole at position $w$. We then plot the quasihole propagators for the Laughlin, Read-Moore and Read-Rezayi states $(k,r)=(1,2),(1,4),(2,2),(3,2)$, which are all described by unitary CFT's. In addition, we plot the quasihole propagator for the Gaffnian \cite{simon2006} state $(k,r)=(2,3)$ which is described by a non-unitary CFT. We find that the $(k,r)=(2,3)$ state has an propagator exponent that seems to be bound by the exponents of the Read-Rezayi and Moore-Read states $\frac{3}{8}< g_{qh} < \frac{3}{10}$ }\label{qhprop}
\end{figure}

We now turn to the quasihole exponent $g_{qh}$. The non-Abelian, fractionalized quasihole Green's function scales as $G(x,t) \sim 1/(x- vt)^{g_{qh}}$, where $g_{qh}$ is usually less than unity. As we later prove, most of the FQH Jack polynomials are related to non-unitary CFT's, which have unphysical negative quasihole exponents $g_{qh}<0$. Our aim is hence not to compute the exact value of $g_{qh}$, but to prove that it is well defined and positive. In particular, we want to compute the equal time non-Abelian quasihole propagator on the edge of the $N$-particle $(k,r)$ FQH state $R= \sqrt{2 r N/k}$ using the quantum mechanical measure on the disk $d \mu_i = \exp(- |z_i|^2/2) d^2 z_i$:
\begin{equation}
G_{qh}(R,R e^{I \phi}) = \frac{\int \prod_{i=1}^N d \mu_i
\psi^\star(R;z_1..z_N) \psi(R e^{i
\phi};z_1..z_N)}{\int \prod_{i=1}^N d \mu_i
\psi^\star(R;z_1..z_N) \psi(R;z_1..z_N)} \nonumber
\end{equation}
\noindent where $\psi(w; z_1..z_N)$ is the wavefunction for the pinned non-Abelian quasihole at position $w$. The physical operation we can perform on a FQH ground-state is the addition of one unit of flux, which corresponds, in this case, to the addition of $k$ charge $\frac{1}{r}$ quasiholes, which then further fractionalize and become distinct.  We then take $k-1$ quasiholes to the origin (in the disk geometry) and the $k$'th
non-Abelian quasihole at $w$ (see Fig[\ref{qhprop}]). This fractionalized quasihole wavefunction $\psi(w; z_1..z_N)$ is defined by the following clustering property:
\begin{equation}
\psi(w; z_1,..., z_N)|_{z_1 =...=z_k =w} =0
\end{equation}
\noindent which destroys the wavefunction if $k$ particles reach the fractionalized quasihole $w$. A further condition $\psi(w; z_1,..., z_N)|_{z_1 =z_2=0} =0$ pins $k-1$ quasiholes at the origin.
 A nice expression for $\psi(w; z_1..z_N)$ is currently known only for the case of the Laughlin and Read-Moore states. For the Laughlin $\nu=\frac{1}{r}$ case, the $1$-quasihole wavefunction reads:
\begin{equation}\label{laughlinquasihole}
\psi = \prod_{i=1}^N (z_i- w) \prod_{i<j}^N (z_i-z_j)^r = \sum_{i=0}^N (-w)^i P_i(z_1,...,z_N) \nonumber
\end{equation}
\noindent where the second expression is the coherent state representation of the pinned quasihole wavefunction, with $P_i(z_1,...,z_N) = m_{1^i}(z_1,...,z_N) \prod_{i<j}^N (z_i-z_j)^r$, and $m_{1^i}$ is the monomial symmetric function. We found the generalization of the above expression to all non-Abelian states. We begin by defining the following Jack polynomials in electron coordinates (we use the "monic" Jack normalization, where the coefficient of the monomial of root partition in the Jacks is always unity):
\begin{equation}
\begin{array}{cc}
  |0\rangle  = & J^{\alpha(k,r)}_{0k0^{r-1}k0^{r-1}k ... 0^{r-1} k0^{r-1} k}  \\
  |1\rangle = &   J^{\alpha(k,r)}_{1k-10^{r-1}k0^{r-1}k ...0^{r-1} k0^{r-1} k} \\
  |2 \rangle =  &    J^{\alpha(k,r)}_{1k-10^{r-2}1k-10^{r-1}k ...0^{r-1}k 0^{r-1} k}\\
 |3 \rangle  = &    J^{\alpha(k,r)}_{1k-10^{r-2}1k-10^{r-2}1k-1 ...0^{r-1}k 0^{r-1} k}\\
  \vdots &  \vdots   \\
  |\frac{N}{k} -1 \rangle = &   J^{\alpha(k,r)}_{1k0^{r-2}1k-10^{r-2}1k-1 ... 0^{r-2}1 k-1 0^{r-1} k} \\
  |\frac{N}{k} \rangle = &  J^{\alpha(k,r)}_{1k-10^{r-2}1k-10^{r-2}1k-1 ...0^{r-2}1 k-1 0^{r-2} 1k-1} \\
\end{array}\label{jackquasihole}
\end{equation}
We then find the wavefunction for $k-1$ quasiholes at the origin, $1$-non-Abelian quasihole at position $w$:
\begin{equation} \label{coherentstate}
\psi(w;z_1,...,z_N)=\sum_{i=0}^{\frac{N}{k}}
\left(-\frac{w}{k}\right)^i |i\rangle
\end{equation}
\noindent For $k=1$ this reduces to the Laughlin case, as the polynomials $P_i(z_1,...,z_N)$ are identical to the Jacks $|i\rangle$ in Eq.(\ref{jackquasihole}) for $(k,r)=(1,r)$. We find that the Jacks in Eq.(\ref{jackquasihole}) are related by the angular momentum operator $L^-= N_\Phi \sum_{i=1}^N z_i  - \sum_{i=1}^N z_i^2 \partial/\partial z_i$:
\begin{equation}
\left(L^-\right)^i |\frac{N}{k} \rangle = i! k^i |\frac{N}{k} - i
\rangle
\end{equation}
After some algebra we obtain:
\begin{equation}
L^- \psi(w;z_1..z_N) = ( w^2\frac{\partial}{\partial
w}- \frac{N}{k} w )\psi(w;z_1..z_N)
\end{equation}
\noindent
Hence, when written as a coherent state representation of Jacks, the non-Abelian quasihole wavefunction takes an identical form for all non-Abelian states. Using this wavefunction, we have numerically computed the quasihole propagator at the edge $R$ of the FQH droplet $|G_{qh}(R,R e^{i \phi})|$ which we plot in Fig.[\ref{qhprop}]. We see that the propagator for the state $(k,r)=(2,3)$, which we pick as an example of a state described by a non-unitary CFT, does not differ, qualitatively, from the propagators of states described by unitary CFTs: it is well behaved, and falls off with the arc distance $\phi$, properties which we have verified are shared by all other Jacks. This is in contrast to what the FQH-CFT connection predicts for non-unitary CFTs: a quasihole exponent which grows with the arc distance.

We now obtain the electron propagator by assuming that our wavefunctions can be written as correlators of primary fields in a CFT. Let the electron propagator be $\psi_e(z)= \psi_1(z) \exp(i \alpha \phi(z))$, with $\alpha$ an unknown boson compactification radius. Any FQH wavefunction, be it the ground-state or the quasihole excitations, must be single-valued in electron coordinates. The clustering property of the $(k,r)$ Jack polynomials gives, for $\{z\}$ $\equiv$
$\{z_1,\ldots,z_N\}$,
\begin{equation}
J^{\alpha_{k,r}}_{\lambda^0(k,r)}(\{z\},Z,\ldots,Z)=\prod_{i=1}^N (Z-z_i)^r J^{\alpha_{k,r}}_{\lambda^0(k,r)}(\{z\})  , \label{jackclustering}
\end{equation}
where on the RHS, $z_i = Z$ for $i$ = $N+1,\ldots, N+k$, and $n(\lambda^0(k,r))=k0^{r-1}k0^{r-1}k..k0^{r-1}k$, the ground-state occupation number.
 Clustering electrons together is equivalent to fusing in the language of CFT, so Eq.(\ref{jackclustering}) expresses the fusion of $k$ electron operators at the point $Z$ which then fuse with the remaining electrons at points $z_1,...,z_N$. The overall Laughlin factor in Eq.(\ref{jackclustering}) arises from the clustering of the bosonic vertex operators $\exp(i \alpha \phi(z))$. To perform the fusion of electron fields around $Z$, let $z_{N+j} = Z+ \epsilon \exp(2\pi i  j/k)$ for $j=1...k$. It takes a trivial calculation to see that a CFT correlator of such field, is equal, up to a constant phase to:
\begin{eqnarray}
&\lim_{\epsilon\rightarrow 0} \langle \prod_{i=1}^N \psi_e(z_i) \prod_{j=1}^K \psi_e(z_{N+j})\rangle = \nonumber \\ &\times \prod_{i=1}^N(Z-z_i)^{k \alpha^2} \frac{\epsilon^{\frac{\alpha^2 k(k-1)}{2}}}{\epsilon^{ k \Delta_{\psi_1}}}  \langle \prod_{i=1}^N \psi_e(z_i)  \psi_{F}(Z)\rangle \label{CFTclustering}
\end{eqnarray}
\noindent where $\Delta_{\psi_1}$ is the scaling dimension of the electron field $\psi_1$ and $\psi_F(Z)$ is the resulting field out of the fusion of $k$ electron fields. By comparing Eq.(\ref{CFTclustering}) with our Jack polynomial clustering Eq.(\ref{jackclustering}), we find that, if we assume our wavefunctions are described by CFT correlators, $\psi_F(Z)=1$ (hence the CFT must have $Z_k$ symmetry, where $k$ fields fuse to the identity), $k \alpha^2=r$ and $k\Delta_{\psi_1} = \alpha^2 k(k-1)/2$. Then $\Delta_{\psi_{1}} =\frac{r(k-1)}{2k}$. The electron propagator exponent is then $2 \Delta_{\psi_1} + \alpha^2 = r$, in accordance to what we have found from our numerical studies.
From the clustering properties of the Jack polynomials and the quasihole wavefunctions (several other quasihole wavefunctions, for different $N$ sectors are needed), we can find the all the fusion rules of all the fields, as well as the scaling dimensions of the Abelian, electron operators \cite{bernevig2008A}. For example, the ground-state Jack polynomial where we fuse $n$ ($\le k$) electrons together at the origin, by taking  $z_j=\epsilon \exp(2 \pi i j/n)$, $j=0...n$ and then $\epsilon \rightarrow 0$ becomes $\lim_{\epsilon \rightarrow 0} J^{\alpha(k,r)}_{k0^{r-1}k0^{r-1}k...k0^{r-1}k} (z_1,...z_n,z_{n+1},...,z_N) =J^{\alpha(k,r)}_{k-n0^{r-1}k0^{r-1}k...k0^{r-1}k} (z_{n+1},...,z_N)$. As the right hand side does not depend on $\epsilon$, the $n$ fields $\psi_1$ have fused into a $\psi_n$ field with scaling dimension $\Delta_n$ which is obtained by cancelling the dependence on $\epsilon$: the fusion of the $\psi_1$ fields into the $\psi_n$ gives a factor of $1/\epsilon^{n \Delta_1 - \Delta_n}$ while the $U(1)$ bosonic factors $\exp(i \alpha \phi(z))$ give a factor $\prod_{i<j=1}^n (z_i-z_j)^{\alpha^2} \sim \epsilon^{n(n-1)\alpha^2/2}$. The denominator must cancel the numerator and we obtain $\Delta_n = n r(k-n)/2k$.

Similar methods allow for the fusion rules and the $U(1)$ charges (boson compactification radiuses) of the quasihole excitations. Let the fractionalized quasihole operator be $\psi_{qh}(z) = \sigma_1(z) \exp(i \beta \phi(z))$. From the expression of our fractionalized quasihole, Eq.(\ref{jackquasihole}), when fusing the remaining quasihole at the origin $w\rightarrow 0$, the boson correlator of the $k$ quasihole fields at the origin fused with $N$ electron fields gives $\prod_{i=1}^N z_i^{k \beta \alpha}$. As only the $|0\rangle$ state remains in Eq.(\ref{jackquasihole}), which is $\prod_{i=1}^N z_i$ times the $N$-electron ground-state, it is immediate that $\beta = 1/(k\alpha) =1/\sqrt{k r}$. The fusion rules, however, are more complicated because the polynomial form of our wavefunction knows only about expectation values of CFT correlators. As such, the full fusion rules can be obtained only when the quasihole wavefunctions are known in the sectors with different fermionic number (for our $(k,r)$ states, there can be $k$ fermionic sectors, which are characterized by $N$, $N-1$,...,$N-k+1$ divisible by $k$). We found that the quasihole wavefunctions in different fermionic sectors can also be obtained as coherent state representations of Jack polynomials. We reserve this and the fusion rules that the quasihole excitations obey for a future publication \cite{bernevig2008A}. The only quantities that cannot be obtained from the polynomial form of the wavefunctions are the quasihole scaling dimensions. The reason for this is that, since the wavefunction in Eq.(\ref{coherentstate}) needs to be single-valued only in the electron coordinates, we can multiply our quasihole wavefunction by any overall function that depends only on the quasihole coordinates to obtain another perfectly acceptable un-normalized wavefunction with the same physics. To compute the quasihole scaling dimension, one needs to compute the quantum-mechanical norm of the Jack polynomials (this would also allow for an exact computation of braiding matrices). We have done this in the Laughlin case exactly \cite{bernevig2008A}, but so far we have not been able to generalize it to non-Abelian states.

The results above strongly support the conjecture, first presented in \cite{feigin2002} that the Jack polynomials are the correlation functions of $W_k(k+1,k+r)$ models (these are generalizations of the $Z_k$ parafermions, see appendix of \cite{zamolodchikov1985}). The thermal Hall coefficient we found from partition counting is equivalent to the effective central charge of the $W_k(k+1,k+r)$ models. The same is true for the electron propagator, electron scaling dimension, and the fusion rules. For $r\ge 3$, the $W$ algebras are non-unitary, and the quasihole field has negative scaling dimension. Using the CFT scalar product to compute quasihole propagators, they blow up with the arc-distance, certainly an unphysical feature. However, if we compute the quasihole exponent with the quantum mechanical scalar product we find it is perfectly well-defined and positive. Hence the CFT-FQH correspondence breaks down for non-unitary CFT's and other methods are needed for computing quasihole correlators. We conjecture that non-unitary CFT's have an "effective" positive scaling dimension (it is known that they have an effective central charge which is different from the sometimes negative central charge). It is certainly true that the negative power of the quasihole correlator is a sickness of the negative scalar product of the non-unitary CFT -- we can also obtain negative scaling dimensions if we use another non-unitary scalar product for Jack polynomials, the combinatoric one \cite{bernevig2008A}.

We wish to thank S. Sondhi, D. Huse, S.Shenker, E.Fradkin and N. Read for discussions and in particular to S. H. Simon for numerous discussions, for a critical reading of the manuscript, and for pointing us to references regarding $W$ algebras. This work was supported in part by the U.S. National Science Foundation (under MRSEC Grant No. DMR-0213706
at the Princeton Center for Complex Materials).


\begin{thebibliography}{23}
\expandafter\ifx\csname natexlab\endcsname\relax\def\natexlab#1{#1}\fi
\expandafter\ifx\csname bibnamefont\endcsname\relax
  \def\bibnamefont#1{#1}\fi
\expandafter\ifx\csname bibfnamefont\endcsname\relax
  \def\bibfnamefont#1{#1}\fi
\expandafter\ifx\csname citenamefont\endcsname\relax
  \def\citenamefont#1{#1}\fi
\expandafter\ifx\csname url\endcsname\relax
  \def\url#1{\texttt{#1}}\fi
\expandafter\ifx\csname urlprefix\endcsname\relax\def\urlprefix{URL }\fi
\providecommand{\bibinfo}[2]{#2}
\providecommand{\eprint}[2][]{\url{#2}}

\bibitem[{\citenamefont{\text{N. Read} and \text{E. Rezayi}}(1999)}]{read1999}
\bibinfo{author}{\bibnamefont{\text{N. Read}}} \bibnamefont{and}
  \bibinfo{author}{\bibnamefont{\text{E. Rezayi}}}, \bibinfo{journal}{Phys.
  Rev. B} \textbf{\bibinfo{volume}{59}}, \bibinfo{pages}{8084}
  (\bibinfo{year}{1999}).

\bibitem[{\citenamefont{\text{A.B. Zamolodchikov} and \text{V.A.
  Fateev}}(1985)}]{zamolodchikov1985}
\bibinfo{author}{\bibnamefont{\text{A.B. Zamolodchikov}}} \bibnamefont{and}
  \bibinfo{author}{\bibnamefont{\text{V.A. Fateev}}}, \bibinfo{journal}{Sov.
  Phys. JETP} \textbf{\bibinfo{volume}{62}}, \bibinfo{pages}{2}
  (\bibinfo{year}{1985}).

\bibitem[{\citenamefont{\text{E. Ardonne}}(2002)}]{ardonne2002}
\bibinfo{author}{\bibnamefont{\text{E. Ardonne}}}, \bibinfo{journal}{J. Phys.
  A: Math. Gen} \textbf{\bibinfo{volume}{35}}, \bibinfo{pages}{447}
  (\bibinfo{year}{2002}).

\bibitem[{\citenamefont{\text{B.A. Bernevig} and \text{F.D.M.
  Haldane}}({\natexlab{a}})}]{bernevig2007}
\bibinfo{author}{\bibnamefont{\text{B.A. Bernevig}}} \bibnamefont{and}
  \bibinfo{author}{\bibnamefont{\text{F.D.M. Haldane}}},
  \bibinfo{howpublished}{arXiv:0707.3637}.

\bibitem[{\citenamefont{\text{N. Read}}()}]{read2007}
\bibinfo{author}{\bibnamefont{\text{N. Read}}},
  \bibinfo{howpublished}{arXiv:0711.0543v1}.

\bibitem[{\citenamefont{\text{S.H. Simon} et~al.}(2007)\citenamefont{\text{S.H.
  Simon}, \text{E.H. Rezayi}, \text{N.R. Cooper}, and \text{I.
  Berdnikov}}}]{simon2006}
\bibinfo{author}{\bibnamefont{\text{S.H. Simon}}},
  \bibinfo{author}{\bibnamefont{\text{E.H. Rezayi}}},
  \bibinfo{author}{\bibnamefont{\text{N.R. Cooper}}}, \bibnamefont{and}
  \bibinfo{author}{\bibnamefont{\text{I. Berdnikov}}}, \bibinfo{journal}{Phys.
  Rev. B} \textbf{\bibinfo{volume}{75}}, \bibinfo{pages}{075317}
  (\bibinfo{year}{2007}).

\bibitem[{\citenamefont{\text{B.A. Bernevig} and \text{F.D.M.
  Haldane}}({\natexlab{b}})}]{bernevig2008}
\bibinfo{author}{\bibnamefont{\text{B.A. Bernevig}}} \bibnamefont{and}
  \bibinfo{author}{\bibnamefont{\text{F.D.M. Haldane}}},
  \bibinfo{howpublished}{in preparation}.

\bibitem[{\citenamefont{\text{B. Feigin} et~al.}(2002)\citenamefont{\text{B.
  Feigin}, \text{M. Jimbo}, \text{T. Miwa}, and \text{E. Mukhin}}}]{feigin2002}
\bibinfo{author}{\bibnamefont{\text{B. Feigin}}},
  \bibinfo{author}{\bibnamefont{\text{M. Jimbo}}},
  \bibinfo{author}{\bibnamefont{\text{T. Miwa}}}, \bibnamefont{and}
  \bibinfo{author}{\bibnamefont{\text{E. Mukhin}}}, \bibinfo{journal}{Int.
  Math. Res. Not.} \textbf{\bibinfo{volume}{23}}, \bibinfo{pages}{1223}
  (\bibinfo{year}{2002}).

\bibitem[{\citenamefont{\text{R.P. Stanley}}(1989)}]{stanley1989}
\bibinfo{author}{\bibnamefont{\text{R.P. Stanley}}},
  \bibinfo{journal}{Adv.Math.} \textbf{\bibinfo{volume}{77}},
  \bibinfo{pages}{76} (\bibinfo{year}{1989}).

\bibitem[{\citenamefont{\text{B. Sutherland}}(1971)}]{sutherland1971}
\bibinfo{author}{\bibnamefont{\text{B. Sutherland}}}, \bibinfo{journal}{Phys.
  Rev. A} \textbf{\bibinfo{volume}{4}}, \bibinfo{pages}{2019}
  (\bibinfo{year}{1971}).

\bibitem[{\citenamefont{\text{R. Tao} and\text{D.J. Thouless}}(1983)}]{tao1983}
\bibinfo{author}{\bibnamefont{\text{R. Tao} and\text{D.J. Thouless}}},
  \bibinfo{journal}{Phys. Rev. B} \textbf{\bibinfo{volume}{28}},
  \bibinfo{pages}{1142} (\bibinfo{year}{1983}).

\bibitem[{\citenamefont{\text{B.A. Bernevig} and \text{F.D.M.
  Haldane}}({\natexlab{c}})}]{bernevig2007B}
\bibinfo{author}{\bibnamefont{\text{B.A. Bernevig}}} \bibnamefont{and}
  \bibinfo{author}{\bibnamefont{\text{F.D.M. Haldane}}},
  \bibinfo{howpublished}{arXiv:0711.3062}.

\bibitem[{\citenamefont{\text{J.L. Cardy}}(1988)}]{cardy1988}
\bibinfo{author}{\bibnamefont{\text{J.L. Cardy}}}, \bibinfo{journal}{Phys. Rev.
  Lett.} \textbf{\bibinfo{volume}{60}}, \bibinfo{pages}{2709}
  (\bibinfo{year}{1988}).

\bibitem[{\citenamefont{\text{G.E. Andrews}}()}]{andrews1998}
\bibinfo{author}{\bibnamefont{\text{G.E. Andrews}}},
  \bibinfo{howpublished}{Theory of Partitions, Cambridge University Press (July
  28, 1998)}.

\bibitem[{\citenamefont{\text{W. Nahm} et~al.}(1993)\citenamefont{\text{W.
  Nahm}, \text{A. Recknagel}, and \text{M. Terhoeven}}}]{nahm1993}
\bibinfo{author}{\bibnamefont{\text{W. Nahm}}},
  \bibinfo{author}{\bibnamefont{\text{A. Recknagel}}}, \bibnamefont{and}
  \bibinfo{author}{\bibnamefont{\text{M. Terhoeven}}}, \bibinfo{journal}{Mod.
  Phys. Lett.} \textbf{\bibinfo{volume}{A8}}, \bibinfo{pages}{1835}
  (\bibinfo{year}{1993}).

\bibitem[{\citenamefont{\text{B. Richmond} and \text{G.
  Szekeres}}(1981)}]{richmond1981}
\bibinfo{author}{\bibnamefont{\text{B. Richmond}}} \bibnamefont{and}
  \bibinfo{author}{\bibnamefont{\text{G. Szekeres}}}, \bibinfo{journal}{Math.
  Soc. Ser.} \textbf{\bibinfo{volume}{A 31}}, \bibinfo{pages}{362}
  (\bibinfo{year}{1981}).

\bibitem[{\citenamefont{\text{A.N. Kirilov}}(1995)}]{kirilov1995}
\bibinfo{author}{\bibnamefont{\text{A.N. Kirilov}}}, \bibinfo{journal}{Prog.
  Theor. Phys. Suppl} \textbf{\bibinfo{volume}{118}}, \bibinfo{pages}{61}
  (\bibinfo{year}{1995}).

\bibitem[{\citenamefont{\text{J.-F. Fortin}
  et~al.}(2005{\natexlab{a}})\citenamefont{\text{J.-F. Fortin}, \text{P.
  Jacob}, and \text{P. Mathieu}}}]{fortin2005}
\bibinfo{author}{\bibnamefont{\text{J.-F. Fortin}}},
  \bibinfo{author}{\bibnamefont{\text{P. Jacob}}}, \bibnamefont{and}
  \bibinfo{author}{\bibnamefont{\text{P. Mathieu}}},
  \bibinfo{journal}{Electronic J. Comb.} \textbf{\bibinfo{volume}{12}},
  \bibinfo{pages}{R12} (\bibinfo{year}{2005}{\natexlab{a}}).

\bibitem[{\citenamefont{\text{B. Feigin} and
  \text{\emph{et.al.}}}()}]{feigin2002B}
\bibinfo{author}{\bibnamefont{\text{B. Feigin}}} \bibnamefont{and}
  \bibinfo{author}{\bibnamefont{\text{\emph{et.al.}}}},
  \bibinfo{howpublished}{arXiv:math/0212347v1}.

\bibitem[{\citenamefont{\text{J.-F. Fortin}
  et~al.}(2005{\natexlab{b}})\citenamefont{\text{J.-F. Fortin}, \text{P.
  Jacob}, and \text{P. Mathieu}}}]{fortin2005B}
\bibinfo{author}{\bibnamefont{\text{J.-F. Fortin}}},
  \bibinfo{author}{\bibnamefont{\text{P. Jacob}}}, \bibnamefont{and}
  \bibinfo{author}{\bibnamefont{\text{P. Mathieu}}},
  \bibinfo{journal}{Ramanujan Journal} \textbf{\bibinfo{volume}{10}},
  \bibinfo{pages}{215} (\bibinfo{year}{2005}{\natexlab{b}}).

\bibitem[{\citenamefont{\text{B. Feigin} et~al.}()\citenamefont{\text{B.
  Feigin}, \text{M. Jimbo}, and \text{T. Miwa}}}]{feigin2002C}
\bibinfo{author}{\bibnamefont{\text{B. Feigin}}},
  \bibinfo{author}{\bibnamefont{\text{M. Jimbo}}}, \bibnamefont{and}
  \bibinfo{author}{\bibnamefont{\text{T. Miwa}}},
  \bibinfo{howpublished}{arXiv:math/0012193v2}.

\bibitem[{\citenamefont{\text{X.-G. Wen}}(1992)}]{wen1992}
\bibinfo{author}{\bibnamefont{\text{X.-G. Wen}}}, \bibinfo{journal}{Int. J.
  Mod. Phys. B} \textbf{\bibinfo{volume}{6}}, \bibinfo{pages}{1711}
  (\bibinfo{year}{1992}).

\bibitem[{\citenamefont{\text{B.A. Bernevig}}()}]{bernevig2008A}
\bibinfo{author}{\bibnamefont{\text{B.A. Bernevig}}}, \bibinfo{howpublished}{in
  preparation}.

\end{thebibliography}
\end{document}